# Phase Instability amid Dimensional Crossover in Artificial Oxide Crystal

Seung Gyo Jeong[1,†] , Taewon Min[2,†], Sungmin Woo[1], Jiwoong Kim[2], Yu-Qiao Zhang[3], Seong Won Cho[4,5], Jaeseok Son[6,7], Young-Min Kim[8,9], Jung Hoon Han[1], Sungkyun Park[2], Hu Young Jeong[10], Hiromichi Ohta[3], Suyoun Lee[4], Tae Won Noh[6,7], Jaekwang Lee[2,‡] and Woo Seok Choi[1,*]

[1]Department of Physics, Sungkyunkwan University, Suwon 16419, Korea

[2]Department of Physics, Pusan National University, Busan 46241, Korea

[3]Research Institute for Electronic Science, Hokkaido University, Sapporo 001-0020, Japan

[4]Electronic Materials Research Center, Korea Institute of Science and Technology, Seoul, 02792, Korea

[5]Department of Materials Science and Engineering, Seoul National University, Seoul, 08826, Korea

[6]Department of Physics and Astronomy, Seoul National University, Seoul 08826, Korea

[7]Center for Correlated Electron Systems, Institute for Basic Science, Seoul 08826, Korea

[8]Department of Energy Sciences, Sungkyunkwan University, Suwon 16419, Korea

[9]Center for Integrated Nanostructure Physics, Institute for Basic Science, Suwon 16419, Korea

[10]UNIST Central Research Facilities and School of Materials Science and Engineering, Ulsan National Institute of Science and Technology, Ulsan 44919, Korea

Artificial crystals synthesized by atomic-scale epitaxy provides the ability to control the dimensions of the quantum phases and associated phase transitions via precise thickness modulation. In particular, reduction in dimensionality via quantized control of atomic layers is a powerful approach to revealing hidden electronic and magnetic phases. Here, we demonstrate a dimensionality-controlled and induced metal-insulator transition (MIT) in atomically designed superlattices by synthesizing a genuine two dimensional (2D) $SrRuO_3$ crystal with highly suppressed charge transfer. The tendency to ferromagnetically align the spins in $SrRuO_3$ layer diminishes in 2D as the interlayer exchange interaction vanishes, accompanying the 2D localization of electrons. Furthermore, electronic and magnetic instabilities in the two $SrRuO_3$ unit cell layers induce a thermally-driven MIT along with a metamagnetic transition.

The metal-insulator transition (MIT) is one of the representative phenomena observed in transition metal oxides [1-3]. $3d$ perovskite oxide systems foster MIT, which is understood in terms of the competition between the itinerant nature of the charge carriers and electronic correlation. For example, (La,Sr)TiO$_3$, (La,Sr)MnO$_3$, and (La,Sr)CoO$_3$ show MIT via charge carrier doping, which alleviates the Mott insulating state. In contrast, SrVO$_3$ and LaNiO$_3$ show a decrease in the bandwidth with decreasing film thickness. On the other hand, MIT is not as common in $4d$ perovskite oxides. Itinerant ferromagnet SrRuO$_3$ (SRO) is one of the $4d$ oxides that exhibits a thickness-dependent MIT. That is, it is known to become insulating as the film thickness decreases below ~4 perovskite unit cells (u.c.) in most cases [4-7]. Experimentally, growth-induced disorder has been predominantly proposed as the origin of the MIT [4,8], while other intrinsic mechanisms, including quantum confinement and orbital ordering, have also been suggested theoretically [9]. In addition, the close relationship between the spin state and electronic structure has been intensively studied [4,7-10]. Yet, a coherent observation and interpretation of the thickness-dependent MIT is lacking, in part because of the absence of a low dimensional sample with intrinsic SRO layers.

In the meantime, recent advances in atomic scale epitaxy have enabled the realization of synthetic crystals with customized dimensions. It is possible to construct a testbed for emergent phenomena in the quasi-two-dimensional system by deliberately designing the atomic structure. Indeed, artificial superlattices (SLs) composed of perovskite oxides provide direct access to low dimensional physical phenomena by manipulating the structural periodicity [11]. Particularly, 2-dimensional electron liquids have been intensively studied for quantum mechanical behavior such as quantum confinement, resonant tunneling, and thermopower enhancement [12-14]. Such unprecedented capability lets us explore the fundamental aspects of MIT in low dimensions.

In this paper, we report an intrinsic MIT in SRO with electronic interfaces as sharp as atomic or chemical interfaces. A close correlation between the electronic structure, transport properties, and spin ordering is revealed, which is strongly dependent on the system dimensions. Thus, the 3 u.c. layer of SRO is ferromagnetic metallic (FM-M), while the 1 u.c. one is antiferromagnetic insulating (AFM-I). Furthermore, a phase instability is observed for the SRO layer with 2 u.c. thickness, which bridges the 2- and 3-dimensional states. Herein, a thermally driven MIT with strongly coupled magnetic ordering is found, manifesting the dimensional instability between the FM-M phase and AFM-I phases. Our results describe the evolution of the strongly coupled electronic and magnetic phases via atomic-scale dimensional engineering of correlated materials.

Customization of the dimension of the SRO crystals was achieved by systematically controlling the number of atomic u.c. layers of SRO and SrTiO$_3$ in [(SRO)$_x$|(STO)$_y$]$_{10}$ ($x$ u.c. layers of SRO and $y$ u.c. layers of STO repeated 10 times along the growth direction, [$x$|$y$]) SLs using pulsed laser deposition (PLD) (Figs. S2-S4 and Supplemental Note 1 [15]). We note that the epitaxial strain can slightly modify

the electromagnetic properties of SRO [18]. However, all the SRO/STO SLs shown here are coherently strained to the STO substrate (Fig. S4(b)), negating the degree of epitaxial strain as a control parameter. The absence of charge transfer across the SRO/STO interface was confirmed from optical spectroscopic results, which suggest that the SRO/STO SLs embody intrinsic SRO layers. Figure 1(a) shows the optical conductivity spectra ($\sigma_1(\omega)$) of the SLs with varying periodicity. The increase in $\sigma_1(\omega)$ below ~1 eV for most of the SLs manifest the Drude contribution, reflecting the metallic nature. Notably, an isosbestic-like point appears at ~3.86 eV, indicating that the optical sum rule is obeyed for the SL systems. The spectral weight is exchanged across the reference point of the isosbestic-like point at which the values of optical conductivity is the same for bare SRO and STO. Indeed, the total optical spectral weight (*SW*) of the SRO/STO SLs changes *linearly* as a function of the $x/(x + y)$ volume ratio of the SRO layer, as shown in Fig. 1(b). Moreover, the *SW*s below and above the isosbestic point of 3.86 eV are linearly exchanged with each other, suggesting that $\sigma_1(\omega)$s of the SLs can be simply constructed from a linear combination of the $\sigma_1(\omega)$s of the individual SRO and STO layers. (Note that the total *SW* increases slightly with $x/(x + y)$ because $\sigma_1(\omega)$ was measured only up to ~5.5 eV, such that the contribution from the higher energy tail is excluded.) The absence of contributions from the interfacial electronic state indicates the suppression of charge transfer across the interface between SRO and STO (Fig. S5 and Supplemental Note 3 [19]).

In general, the electronic density profile at the interface is broader than the chemical interface due to the prevailing charge transfer at the isostructural perovskite oxide heterointerfaces [22]. While polar discontinuity is a prominent example of the electronic reconstruction of the interface between a polar and a non-polar material [23], even some non-polar materials can form an interface with a substantial charge transfer manifesting the strong covalent or ionic bonding in the oxides [24]. Recently, Zhong and Hansmann proposed a band alignment scheme to quantitatively estimate the charge transfer at the complex oxide interfaces [25]. The misalignment of oxygen *p*-bands across the heterointerface can yield a Fermi energy mismatch, which might lead to an electron transfer across the interface. According to their results, a small Fermi energy mismatch between SRO and STO can result in an interface with highly suppressed charge transfer. Our first-principles density functional theory (DFT) calculations support the existence of an atomically-sharp-electronic-interface, as the layered density of states (LDOS) has no defect or interface states associated with charge transfer (Fig. 1(d)). Therefore, a genuinely 2D SRO system is realized through the SRO/STO SLs, and such a 2D system is necessary for the study of the fundamental physical properties of the SRO layers.

The atomically-sharp-electronic-interface pushes the thickness limit of 2D metallicity in SRO down to 2 u.c. The metallicity of SRO was quantitatively characterized based on the effective carrier concentration ($n_{eff}$) obtained from the Drude model of the $\sigma_1(\omega)$ below ~1 eV (Fig. 1(c) and Supplemental Note 4 [26]). The linear behavior of $n_{eff}$ as a function of $x/(x + y)$ suggests the absence of the charge transfer across the interface and once again further indicates that the metallicity of the SRO is well-maintained to $x = 2$ at room temperature. The metallic behavior abruptly disappeared at $x = 1$ SL with an intrinsic gap opening ($E_g \sim 30$ meV) (Fig. 1(a), manifesting the dimensionality-induced MIT.

DC electrical transport confirmed the MIT at $x = 1$ SL (Fig. 2(a), which is consistent with the optical spectra. In contrast to the previous studies [4,5,8], the temperature-dependent electrical resistivity ($\rho(T)$) of the SLs well preserve the intrinsic metallic nature of SRO down to 2 u.c. This is due to the elimination of extrinsic factors such as surface structural/electronic reconstruction and/or disorder in the layers (Figs. S6, S7 [28], and Supplemental Note 5 [33]). The $\rho(T)$ curve shows an insulating behavior only for the $x = 1$ SL. The activation gap was estimated to be ~20 meV (Fig. S8), which is similar to the optical gap. Thermopower ($S(T)$) measurements for the SLs also showed a clear distinction between the metallic ($x \geq 2$) and insulating ($x = 1$) SLs. As shown in Fig. 2(b), a large suppression of hole carriers in the $x = 1$ SL led to a drastic increase in the thermopower, resembling the behavior of conventional semiconductors (Supplemental Note 6 [35]).

As the metallicity disappears at the 2D limit of the SRO layers, the spin-ordered state undergoes a concomitant change. Figure 2(c) shows the field-cooled (FC) magnetization as a function of temperature ($M(T)$). The conventional ferromagnetism of the SRO layers is systematically suppressed, with a decrease in the FM transition temperature $T_c$, as $x$ decreases (inset of Fig. 2(c)) [42]. When $x$ becomes one, the FM behavior completely disappears over the entire temperature range, suggesting a dimensionality-induced transition from a FM to an AFM or paramagnetic state. This magnetic phase transition at $x = 1$ SL coincides with the dimensional crossover of the electronic state observed by optical and electronic transport measurements shown previously. DFT calculations provide further insight into this magnetically coupled MIT (Fig. S9). Figure 2(d) shows the density of states (DOS) of the SLs with $x = 1$, 2, and 3. $x = 1$ and 2 SLs show an AFM-I phase, with a band gap opening between the $d_{xy}$ and $d_{yz,zx}$ of the Ru-$t_{2g}$ orbitals corresponding to the crystal field splitting (Fig. S10) [9]. In contrast, $x = 3$ SL shows an FM-M phase with a closed gap and spin-polarized DOS, wherein the $t_{2g}$ orbitals are more or less uniformly occupied (Fig. S10). The dimensionality-induced magnetic and electronic phase transition can, therefore, be interpreted as an anisotropic hybridization of the Ru-$t_{2g}$ orbitals at low dimension. In particular, the low-dimensional SRO layer breaks the crystalline periodicity at the interfaces by eliminating the Ru-O orbital hybridization along the out-of-plane direction. As a result, the $d_{xy}$ orbital becomes lower in energy than the $d_{yz/zx}$ orbitals, resulting in an effective half-filled system [43]. The highly anisotropic hybridization naturally implies the strongly coupled magnetic and electronic states via anisotropic magnetic exchange interaction, reminiscent of the number-of-layer-dependent spin ordering of some van der Waals layered materials [44]. As schematically shown in Fig. 2(d), the interlayer exchange interaction $J_{inter}$ naturally vanishes for the $x = 1$ SL, stabilizing the AFM-I phase.

The theoretical calculation correctly captures the qualitative nature of the dimensionality-induced MIT strongly coupled to the spin ordering, yet, an apparent discrepancy seems to exist. For the DFT calculation, the MIT occurs at $x = 2$, whereas it occurs at $x = 1$ for the experiment. However, a closer inspection of the $x = 2$ SL shows that the theoretical calculation turns out to be in excellent agreement with the experiment, especially considering that the DFT calculation is performed at 0 K. Specifically, the $x$-dependent total energy difference (Fig. 2(e)) shows that the AFM and FM configurations have nearly degenerate energy for $x = 2$ SL (Fig. S9). The AFM configuration is only ~5 meV/f.u. lower in energy than the FM one, reflecting the phase instability. Thus, one might experimentally expect an AFM-I ground state when the temperature is lowered.

Figure 3 shows an intriguing temperature-dependent magnetic instability for the $x = 2$ SL, capturing the phase instability amid the dimensional crossover. While the $M(T)$ curve of the $x = 2$ SL shows the usual, suppressed $T_c$ of ~75 K due to the decreased dimension in the FC measurement, a broad first-order-like phase transition was observed around 40 K ($T^*$) in the zero-field-cooled (ZFC) measurement (Fig. 3(a)). Figures 3(b) and 3(c) further show that an intriguing temperature-dependent MIT occurred below ~40 K ($T_{MIT}$), with a clear upturn in $\rho(T)$ accompanied by an anomaly in $S(T)$. $T_{MIT}$ synchronized with $T^*$ of the magnetic phase transition for $x = 2$ SL, indicating that the magnetic and electronic degrees of freedom are strongly coupled. Note that this temperature scale also coincides with the total energy difference obtained between the AFM and FM configuration from the DFT calculation, which can be converted to a temperature of ~25.8 K, close to the $T^*$ and $T_{MIT}$ (Supplemental Note 7 [45]).

The nature of the phase transition for $x = 2$ SL can be better perceived from the magnetic-field dependence as shown in Fig. S11 and the inset of Fig. 3(d). Whereas the $M(H)$ curves below $T_c$ show typical ferromagnetic hysteresis loops for the SRO thin film, they evolve into a double hysteresis loop as the temperature is further lowered below $T^*$. The possible spin configurations associated with the appearance of the double hysteresis loop are: interlayer exchange coupling [47], FM/AFM [48] or FM/FM [49] heterostructures with different coercivity, structural distortions, and metamagnetic transition [50-54]. We first rule out the interlayer exchange coupling for the $x = 2$ SL. While the superlattice structure can give rise to the interlayer exchange coupling between the SRO layers separated by the STO layer, we observe similar double hysteresis even in the [2|18] SL with a large enough STO spacer thickness (Fig. S12). Second, double hysteresis can be observed in FM/AFM or FM/FM heterostructures. However, exchange bias should always appear in such magnetic configurations [53], but it is absent in the hysteresis loops of the [2|$y$] SLs. Third, we used structural analyses (Fig. S4(c)) to confirm the absence of anomalous structural distortion in [2|$y$] SLs, which could result in a double hysteresis loop in $M(H)$. The last candidate, i.e., the metamagnetic transition, is more probable. Indeed, a double hysteresis can originate from a field-induced first order metamagnetic transition above a finite magnetic field ($> H^*$) at low temperature ($< T^*$) [52,54]. Magnetoresistance (MR) measurements support the metamagnetic transition based on the closely coupled electronic phase and spin ordering in $x = 2$ SL (Fig. S13). A typical MR behavior of SRO shows a sharp peak at the coercive field ($H_c$) with a negative MR above $H_c$ [55]. However, in the $x = 2$ SL, $\rho(H)$ shows a rather broad maximum below $H^*$ as shown in Fig. S13, indicating an AFM-I phase (Supplemental Note 8). $\rho(H)$ decreased with increasing $H$ only when the FM-M phase was above $H^*$.

The metamagnetic transition in the $x = 2$ SL is summarized in Fig. 3(d) as a function of $H$ and $T$. A similar magnetic transition accompanying the MIT was observed in manganites with colossal magnetoresistance [54], but the origin is fundamentally different from the SRO systems without any charge ordering. Instead, the phase instability in the $x = 2$ SL is achieved by synthetic dimensions. It is also interesting to compare $x = 2$ SL with the natural crystal of layered perovskite $Sr_3Ru_2O_7$, since both possess a close structural similarity considering the two $RuO_6$ octahedral layers. While $Sr_3Ru_2O_7$ does not show an MIT, it does undergo a metamagnetic transition. The metamagnetic transition occurs at a higher magnetic field ($> 7$ T), possibly owing to the different strain state on the SRO layers or chemical environment. Despite these distinctions, it is likely that the dimensional instability plays an important role in inducing the metamagnetic transitions in both materials. A temperature-thickness phase diagram was constructed to synopsize the phases of the low-dimensional SRO (Fig. 3(e)). It summarizes how the typical FM-M evolves into AFM-I phase in the 2D limit and indicates the dimensional instability in between.

In conclusion, we have observed the intrinsic metal-insulator transition (MIT) of SRO via the dimensional crossover in atomically designed SRO/STO SLs. The deliberate design of the SLs suppresses the charge transfer at the interface between SRO and STO, leading to chemically-sharp-electronic-interfaces, thereby providing an intrinsic SRO layer. A clear dimensionality-induced MIT was observed, strongly coupled to the magnetic ordering. For the 2 u.c. SRO layers, a phase amid the dimensional crossover was discovered by observing a temperature-dependent electronic and magnetic phase transition. The result underscores the dimensional instability which can be further extended to general strongly correlated systems.

† S.G.J. and T.M. contributed equally to this work. We are grateful for an insightful discussion with Lingfei Wang, Ambrose Seo, and Jong Mok Ok. This work was supported by the Basic Science Research Programs through the National Research Foundation of Korea (NRF) (NRF-2019R1A2B5B02004546). J.L. acknowledges support from the Samsung Research Funding & Incubation Center of Samsung Electronics under Project Number SRFC-MA1702-01. S.W.C and S.L. were financially supported by Korea Institute of Science and Technology through 2E29570. Y.-M.K. acknowledges financial support provided by the Institute for Basic Science (IBS-R011-D1) and Creative Materials Discovery Program (NRF-2015M3D1A1070672). J.H.H. was supported by the

Samsung Science and Technology Foundation under Project Number SSTF-BA1701-07. H.O. was supported by Grants-in-Aid for Scientific Research A (17H01314) from the Japan Society for the Promotion of Science (JSPS). This work was performed under the Cooperative Research Program of the "Network Joint Research Center for Materials and Devices: Dynamic Alliance for Open Innovation Bridging Human, Environment and Materials".

‡ jaekwangl@pusan.ac.kr,  * choiws@skku.edu

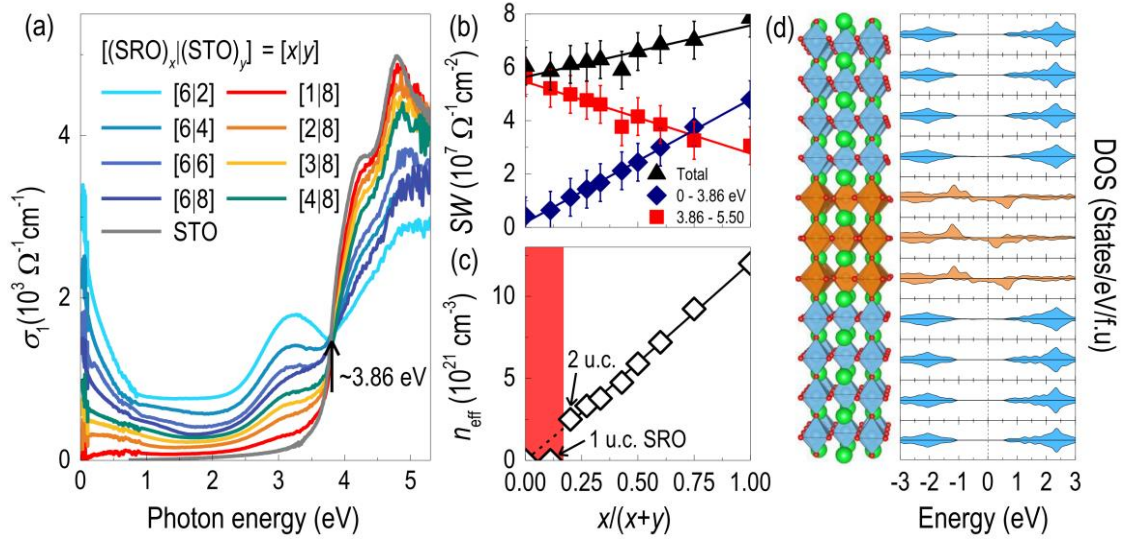

FIG. 1 (color online). Absence of electronic reconstruction in SRO/STO SLs. (a) $\sigma_1(\omega)$ of $[(SRO)_x|(STO)_y]_{10}$ ($[x|y]$) SLs. The ratio between the thicknesses of the SRO and STO layers systematically influence $\sigma_1(\omega)$ of the SLs, with the appearance of an isosbestic-like point at ~3.86 eV. (b) The $W_s$ of the low energy region (0.04 – 3.86 eV, blue rhombus) increases linearly with increasing volume (thickness) the fraction of the SRO layer, whereas the $W_s$ of the high energy region (3.86 – 5.5 eV, red square) decreases linearly. The total $W_s$ is indicated by the black triangles. (c) Effective carrier concentration ($n_{eff}$) of SLs calculated from the Drude contribution of $\sigma_1(\omega)$. A clear MIT is observed for the $x = 1$ SL with an abrupt disappearance of the Drude peak in $\sigma_1(\omega)$ and the resulting $n_{eff}$. The regions in red indicate the insulating phase. (d) The layered density of states (LDOS) of the $[3|8]$ SL does not show any defect or interface states, indicating the absence of the electronic reconstruction. Blue (orange) colored layers are the LDOS from the STO (SRO) layers. The vertical dashed line indicates the Fermi energy level.

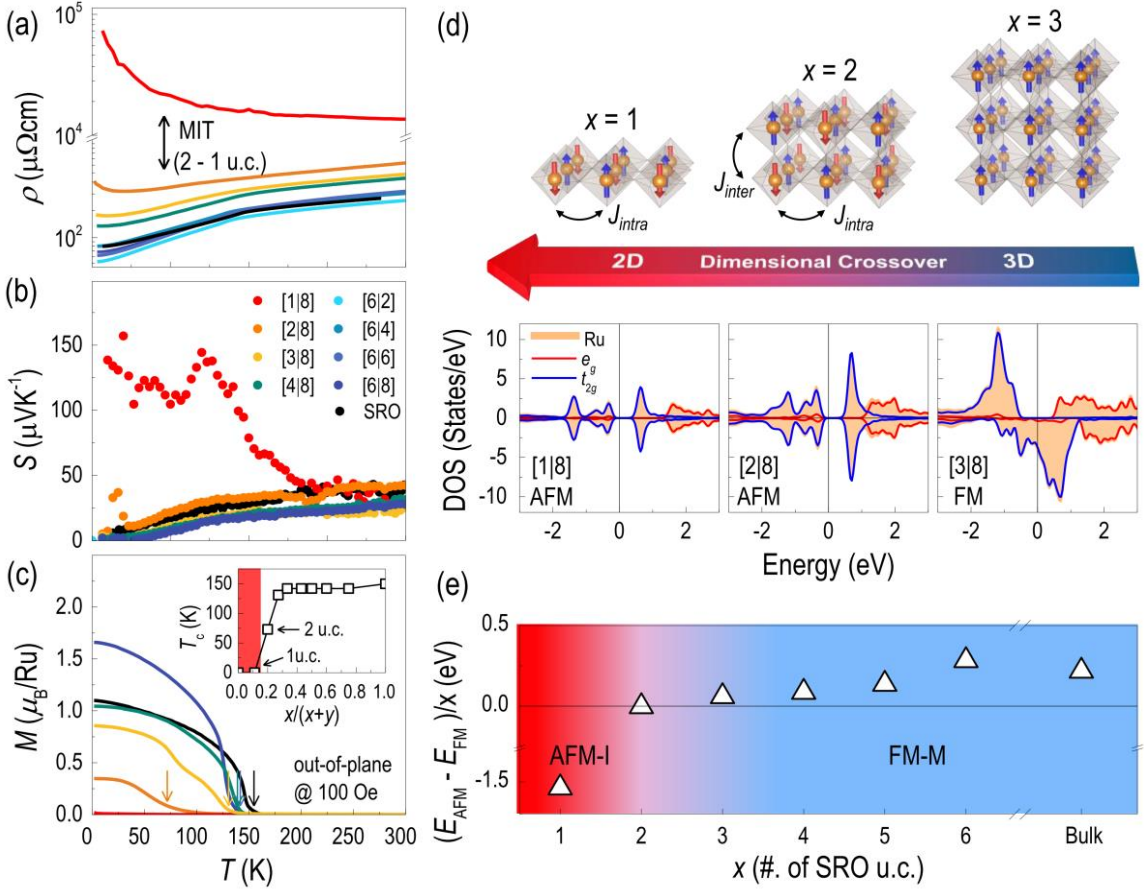

FIG. 2 (color online). Dimensional crossover of electrical transport and magnetic properties. (a) $\rho(T)$ of the $[x|y]$ SLs maintain their metallic behavior down to $x = 2$. A clear MIT is observed for the $x = 1$ SL, consistent with the optical results. (b) $S(T)$ of the SLs shows hole-dominant electrical transport originating from the SRO layers. The $x = 1$ SL shows significantly enhanced $S(T)$ values resembling the behavior of semiconductors with bipolar conduction. (c) $M(T)$ are shown for the SLs, indicating ferromagnetic transitions at low temperature. The arrows indicate the FM $T_c$ of the SLs. The inset shows $T_c$ as a function of SRO fraction within the SLs. Ferromagnetic (FM) ordering of the SRO vanishes for the $x = 1$ SL, concomitant with the MIT. The $M(T)$ curves were measured under 100 Oe of the magnetic field along the out-of-plane direction of the thin films (d) Upper panel: Schematic representation of the spin ordering transition of SRO across the dimensional crossover from 3D to 2D. $J_{intra}$ and $J_{inter}$ represent magnetic exchange coupling along and across the layers, respectively. Lower panel: DOS obtained by DFT calculations of $[x|8]$ ($x = 1, 2,$ and 3) SLs, reproducing the experimentally observed magnetically-coupled MIT. The magnetic phase transition from FM-M ($x = 3$) to AFM-I ($x = 1$) is shown, with the appearance of dimensional instability for the $x = 2$ SL. (e) The calculated energy difference between AFM and FM magnetic configurations as a function of SRO atomic u.c. thickness, where the energy difference is normalized by the SRO thickness. The red (blue) region indicates the AFM-I (FM-M) phase.

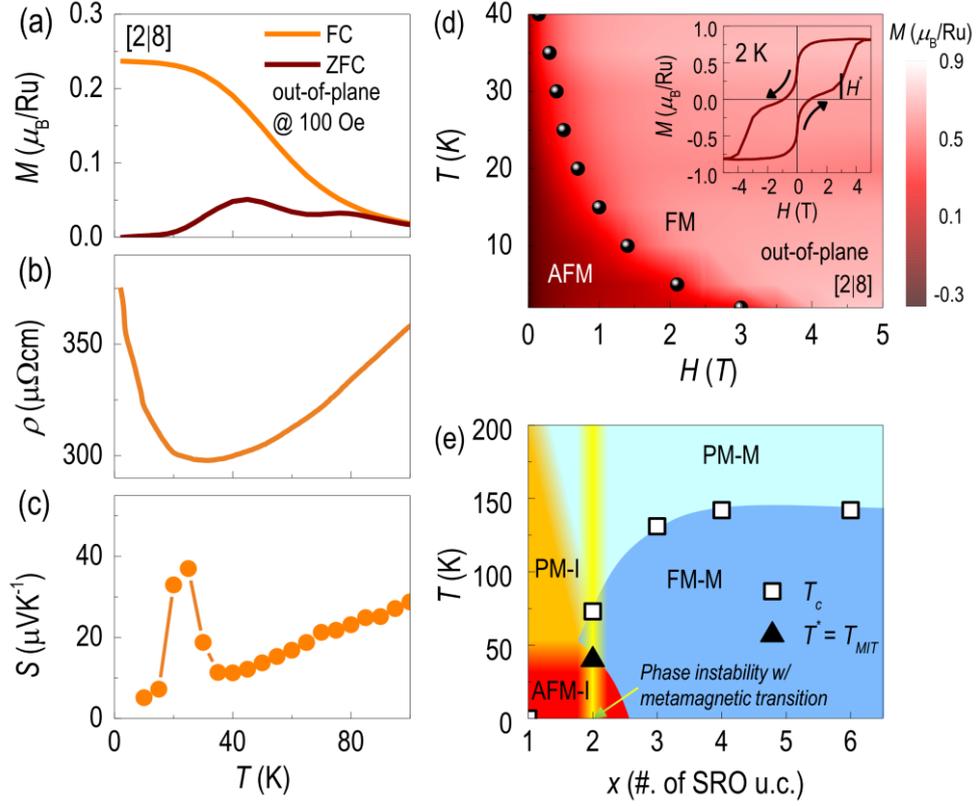

FIG. 3 (color online). Dimensional phase evolution and instability of the atomically designed SRO/STO SLs. (a-c) Dimensional instability in $x = 2$ SL is evidenced by a thermally-induced phase transition occurring simultaneously for the electronic and magnetic structures at ~40 K ($T_{MIT}$). (a) Although FC $M(T)$ shows a single ferromagnetic transition at ~75 K, ZFC $M(T)$ shows a secondary magnetic transition. We applied 100 Oe of the magnetic field along the out-of-plane direction of the thin film. (b) $\rho(T)$ and (c) $S(T)$ indicate that this magnetic transition accompanies the transition to an insulating phase below the $T_{MIT}$. (d) A temperature-magnetic field phase diagram of $x = 2$ SL was plotted from $M(H)$ curves. The data points indicate the magnetic field ($H^*$) required for the metamagnetic transition at different temperatures. The inset shows the magnetic hysteresis behavior of $x = 2$ SL at 2 K. The arrows of the inset indicate the field directions. (e) Phase diagram as a function of SRO thickness and temperature constructed from the experimental and theoretical results, highlighting the dimensionality-induced phase transition and phase instability across the dimensional crossover. The empty squares and solid triangles indicate the $T_c$ and $T^* = T_{MIT}$, respectively. The light blue, blue, yellow and red regions denote PM-M, FM-M, PM-I and AFM-I phases, respectively.

# Phase Instability amid Dimensional Crossover in Artificial Oxide Crystal

Seung Gyo Jeong[1,†] , Taewon Min[2,†], Sungmin Woo[1], Jiwoong Kim[2], Yu-Qiao Zhang[3], Seong Won Cho[4,5], Jaeseok Son[6,7], Young-Min Kim[8,9], Jung Hoon Han[1], Sungkyun Park[2], Hu Young Jeong[10], Hiromichi Ohta[3], Suyoun Lee[4], Tae Won Noh[6,7], Jaekwang Lee[2,‡] and Woo Seok Choi[1,*]

[1]*Department of Physics, Sungkyunkwan University, Suwon 16419, Korea*

[2]*Department of Physics, Pusan National University, Busan 46241, Korea*

[3]*Research Institute for Electronic Science, Hokkaido University, Sapporo 001-0020, Japan*

[4]*Electronic Materials Research Center, Korea Institute of Science and Technology, Seoul, 02792, Korea*

[5]*Department of Materials Science and Engineering, Seoul National University, Seoul, 08826, Korea*

[6]*Department of Physics and Astronomy, Seoul National University, Seoul 08826, Korea*

[7]*Center for Correlated Electron Systems, Institute for Basic Science, Seoul 08826, Korea*

[8]*Department of Energy Sciences, Sungkyunkwan University, Suwon 16419, Korea*

[9]*Center for Integrated Nanostructure Physics, Institute for Basic Science, Suwon 16419, Korea*

[10]*UNIST Central Research Facilities and School of Materials Science and Engineering, Ulsan National Institute of Science and Technology, Ulsan 44919, Korea*

**Supplemental Note 1. Methods.**

**Thin film growth**

[(SrRuO$_3$)$_x$|(SrTiO$_3$)$_y$] superlattices ([(SRO)$_x$|(STO)$_y$]SLs) with controlled $x$ and $y$ u.c. were grown using pulsed laser deposition on (001) STO substrates. Both SRO and STO layers were deposited at 750 °C and 100 mTorr of oxygen partial pressure from the stoichiometric ceramic target using a KrF laser (248 nm; PM864, Lightmachinery). We used a laser fluence of 1.5 J/cm$^2$ and a repetition rate of 5 Hz. Conventional Reflection High Energy Electron Diffraction was not suitable for thickness monitoring because of the high oxygen partial pressure necessary for the stoichiometric film growth. Therefore, we systematically controlled the atomic u.c. of the layers using a customized automatic laser pulse control system programmed in LabVIEW.

**Lattice structure characterization**

X-ray diffraction (XRD) was performed using high-resolution XRD (Rigaku Smartlab and PANalytical X'Pert). Atomic-scale imaging of [(SRO)$_2$|(STO)$_6$] SL were performed on a spherical aberration-corrected scanning transmission electron microscope (STEM; ARM200CF, JEOL) working at 200 kV. To detect weakly scattered signals from oxygen in the structure, annular bright field (ABF) imaging mode was employed along with high-angle annular dark field (HAADF) imaging mode. The incident electron probe angle was 23 mrad, giving rise to a probe size of 0.78 Å. The ABF and HAADF signals were simultaneously collected over detector angle ranges of 7.5 – 17 and 70 – 175 mrad, respectively. Atom-resolved energy dispersive X-ray spectroscopy (EDX) mapping of the SL sample was performed on the same HAADF-STEM imaging mode with a dual EDX detector whose effective X-ray sensing area for each part was 100 mm$^2$. Cross-sectional thin samples for STEM analysis were prepared by a dual-beam focused ion beam system (FIB, FEI Helios Nano Lab 450) and the following low-energy Ar ion milling at 700 V (Fischione Model 1040, Nanomill) was carried out for 15 min to remove damaged surface layers due to heavy Ga ion beam milling in the FIB system.

**Optical property measurements**

The optical properties of the SLs were investigated using spectroscopic ellipsometers (VASE and IR-VASE MARK-2, J. A. Woollam Co.) at room temperature. Optical spectroscopy is the most effective tool for the measurement of intrinsic properties of the embedded layers within a superlattice. The optical spectra ($\sigma_1(\omega)$) were obtained between 0.04 and 5.5 eV for incident angles of 60, 70 and 80°. A two-layer model (SL layer on STO substrate) was used to obtain physically reasonable dielectric functions of the SL system because the wavelength of the probing light was much larger than the u.c. length of the supercell.

**Electrical resistivity, thermopower, and magnetization measurements**

The temperature-dependent electrical resistivity ($\rho(T)$) was measured from 300 to 5 K using the Van der Pauw method with a current source (Keithley 6221) and nanovoltmeter (Keithley 2182) in a commercial cryostat (C-Mag. Vari. 9, Cryomagnetics Inc.). The electrical contacts were formed by indium after scratching the corners of the samples, which give reliable contacts to the buried SRO layers with low contact resistance. The $T$-dependent thermopower ($S(T)$) was measured by the conventional steady-state method within a temperature range of 10 to 300 K by creating a temperature difference ($\Delta T$) of ~2 K across the film. The actual temperatures of both sides of the SRO/STO SLs surface were monitored by using two AuFe 0.07%-chromel thermocouples. The thermo-electromotive force ($\Delta V$) and $\Delta T$ were measured simultaneously, and $S$ values were obtained from the slope of the $\Delta V$–$\Delta T$ plots. $T$- and magnetic field-dependent magnetization ($M(T)$ and $M(H)$) was measured using a Magnetic Property Measurement System (MPMS, Quantum Design). The measurements were performed from 300 to 2 K under 100 Oe of the magnetic field along the out-of-plane direction of the thin films.

## DFT calculations

Our first-principles DFT calculations were performed using the generalized gradient approximation (GGA) and the projector-augmented wave method with a plane-wave basis [S1], as implemented in the Vienna ab-initio simulation package (VASP) code [S2]. Plane waves were included up to the kinetic-energy cutoff 500 eV. For the Brillouin-zone integration, a $\Gamma$-centered $8 \times 8 \times 2$ $k$-point mesh was used for [(SRO)$_x$|(STO)$_8$] SLs and SrO-terminated symmetric (SRO)$_x$ slab model with a vacuum thickness of 16 Å. The in-plane lattice constant of the tetragonal (SRO)$_x$ slabs was fixed to $\sqrt{2}a_{STO}$ with $a^0a^0c^-$ octahedral rotation to realize an antiferromagnetic spin configuration. The on-site Coulomb interaction ($U$) was applied to the Ru-$d$ orbital for [(SRO)$_x$|(STO)$_8$] SLs and (SRO)$_x$ slab calculations. The calculations were converged in energy to $10^{-6}$ eV/cell, and the structures were allowed to relax until the forces were less than $10^{-2}$ eV/Å. In order to obtain the suitable $U$ value, we calculated the magnetic moment and density of states for bulk SRO with increasing $U$ (see Fig. S1). $U$ of 1.6 eV is appropriate for describing the half metallic character and magnetization of bulk SRO.

**Supplemental Note 2. Atomically designed SRO/STO SLs.**

Atomically well-defined SLs with sharp surfaces and interfaces were obtained, as demonstrated by scanning transmission electron microscopy (STEM) images of [(SRO)$_2$|(STO)$_6$] SL (Figs. S2(a), (b)), which were taken using high-angle annular dark field (HAADF) and annular bright field (ABF) imaging modes simultaneously. An energy dispersive X-ray spectroscopy (EDX) map obtained at the atomic scale (Fig. S3) shows chemically well-defined SLs, and the atomic force microscopy results show the atomic step-and-terrace structure, indicating that the surface morphology of the STO substrate is preserved even after the SL growth (inset of Fig. S2(a)). High-resolution X-ray diffraction (HRXRD) $\theta$-$2\theta$ scan in Fig. S2(c) and reciprocal space maps (Fig. S4) also show multiple satellite peaks originating from the periodicity of designed SLs, manifesting the macroscopic ordering of the structure and fully strained nature of the SLs.

**Supplemental Note 3. Effective medium approximation.**

The effective medium approximation (EMA) can be used to derive the average optical response of inhomogeneous composite materials in terms of the volume fractions and macroscopic dielectric properties of the individual materials [S3]. As shown in Fig. S5(b), $\sigma_1(\omega)$ of the SLs could be well-reproduced using the EMA. On the other hand, the EMA could not reproduce $\sigma_1(\omega)$ of the solid solution epitaxial thin film, indicating an electronic reconstruction or charge transfer. $\sigma_1(\omega)$ of the solid solution thin films show distinctive behavior when compared with the SLs with the same SRO ratio ($x$ / $x + y$). Note that this optical result is validated as our optical spectra are consistent with those from the literature on SrRu$_x$Ti$_{1-x}$O$_3$ solid solutions [S4] (Fig. S5(c)). Here, we note that a 2D network of SRO layers with sharp chemical interfaces within the SL is essential for suppressing the charge transfer across the SRO/STO interface and realizing an intrinsic 2D network.

**Supplemental Note 4. Obtaining effective carrier density from $\sigma_1(\omega)$.**

We used the Drude model to extract the effective carrier density ($n_{eff}$) from $\sigma_1(\omega)$ of the SLs. In particular, we used

$$n_{eff} = \frac{e^2}{\omega_p^2 \varepsilon_0 m^*}, \tag{1}$$

where $e$, $\omega_p$, $\varepsilon_0$, and $m^*$ are the electronic charge, plasma frequency, permittivity, and effective mass of carriers. We applied the effective mass of ~4$m_e$ obtained from the de Haas-van Alphen measurement [S5].

**Supplemental Note 5. Thickness dependent resistivity.**

An increase in electric resistivity ($\rho$) is observed below 4 u.c., which can be well understood by assuming an enhanced scattering rate for the extremely narrow SRO channels. Yet, $\rho$(20 K) of the SLs show globally lower resistivity compared to those of the thin films, indicating the absence of possible surface depletion. Assuming $\tau \sim x / v_F$, the thickness- ($x$-) dependent $\rho$ of SRO can be estimated by [S6]

$$\rho = \rho_b + \left(\frac{m^* v_F}{ne^2}\right)\frac{1}{x}, \tag{2}$$

where $\rho_b$, $m^*$, $v_F$, $n$, $e$, and $\tau$ are the bulk resistivity, effective mass, Fermi velocity, carrier density, electronic charge, and scattering time, respectively. The theoretical solid line was drawn using the reported values [8] of $\rho_b \sim 20$ μΩcm, $m^* \sim 4m_e$, $v_F \sim 2 \times 10$ cm/s, $n \sim 1.2 \times 10^{22}$ /cm$^3$, and $\tau \sim 5 \times 10^{-15}$ sec, as shown in Fig. S6. We also summarized $\rho(T)$ of the SRO thin films and other SRO based SLs found in the literature. $\rho(T)$ of SRO for the 2 u.c. shows a better conductivity compared to other references as shown in Fig. S6.

**Supplemental Note 6. Thermopower analysis of MIT.**

We performed the thermopower measurements for investigating the MIT of SRO in Fig. 2(b). The thermopower $S$ of metals can be described in terms of Mott's equation as

$$S = \frac{\pi^2}{3}\left(\frac{k_B^2 T}{3e}\right)\left(\frac{d\ln\sigma(\varepsilon)}{d\varepsilon}\right), \tag{3}$$

where $k_B$, $e$, $\varepsilon$, and $\sigma(\varepsilon)$ are the Boltzmann constant, electronic charge, energy, and electronic conductivity, respectively [S7]. Meanwhile, thermopower $S$ of semiconductors can be expressed as

$$S = \frac{k_B}{e}\left(\frac{E_F - E_v}{k_B T} + A\right), \tag{4}$$

assuming only the hole carrier contribution in thermopower. The $E_F$ and $E_v$ represent Fermi energy and valence band energy, respectively. We can simplify the terms from the relation $n_h = N_v e^{\frac{E_F - E_v}{k_B T}}$, leading to

$$S = \frac{k_B}{e}\left(ln\frac{N_v}{n_h} + A\right). \tag{5}$$

Here, $N_v$ and $n_h$ are the effective density of states of the valence band and the hole carrier density, respectively. $S$ values of metals are typically small and decrease with decreasing temperature. On the other hand, $S$ values of semiconductors increase with decreasing temperature and sensitively depend on $n_h$. Because of this characteristic $S(T)$ behavior between metals and semiconductors, thermoelectric measurements have been adopted to evidence MITs of different bulk [S8, S9] and thin film [S10, S11] materials. However, a semiconducting $S$ behavior has not been observed in the SRO system so far even with various substitutions and doping [S12]. Thus, the demonstration of the drastic change in $S$ with $x$ values in SRO shown in Fig. 2(b) indicates an intrinsic dimension-induced MIT.

**Supplemental Note 7. Néel temperature calculation for 2 u.c. of SRO.**

We calculated the exchange constant ($J$) from the energy difference between AFM and FM spin orderings as follows [S13], assuming that $J_{intra}$ and $J_{inter}$ are similar,

$$J = (E_{AFM} - E_{FM})/N_{nn}N_{mag} = 0.22 \text{ meV}, \tag{6}$$

where $N_{nn}$ and $N_{mag}$ are the number of nearest-neighbor Ru ions and magnetic Ru ions in the $\sqrt{2} \times \sqrt{2} \times 2$ unit cell, respectively. Furthermore, according to the mean-field approximation, the *Né*el temperature is given by as follows:

$$T_N = J \frac{S(S+1)}{3k_B} N_{nn} = 25.8 \text{ K}, \tag{7}$$

where $S$ is the total spin of Ru ion and $k_B$ is the Boltzmann constant. The result shows a very good agreement with the experimentally measured $T^*$ for $x = 2$ SL.

**Supplemental Note 8. Magnetic ground state of $x = 2$ SL.**

The magnetic state of $x = 2$ SL at low-$T$ and low-$H$ can either be AFM or paramagnetic, but the AFM-I phase is more plausible and consistent with the DFT calculation. Particularly, a phase transition from a spin ordered state to a disordered state with decreasing temperature (entropy) is thermodynamically infeasible. In addition, ZFC-$M(T)$ does not follow Curie's law below $T^*$, as shown in Fig. 2(a). On the other hand, a ferrimagnetic ground state is also not viable, since it is highly likely that the spin moments of the Ru ions are identical as they sit on identical atomic sites.

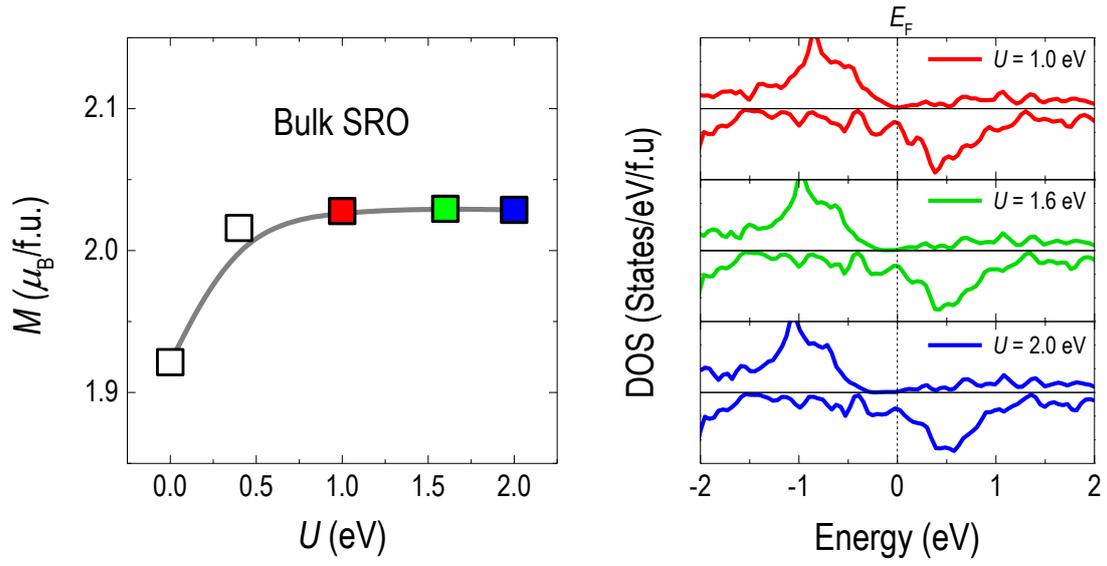

FIG. S1. The magnetic moment of bulk SRO as a function of $U$ (left panel), and the electronic DOS of bulk SRO for $U = 1.0$, 1.6, and 2.0 eV (right panel).

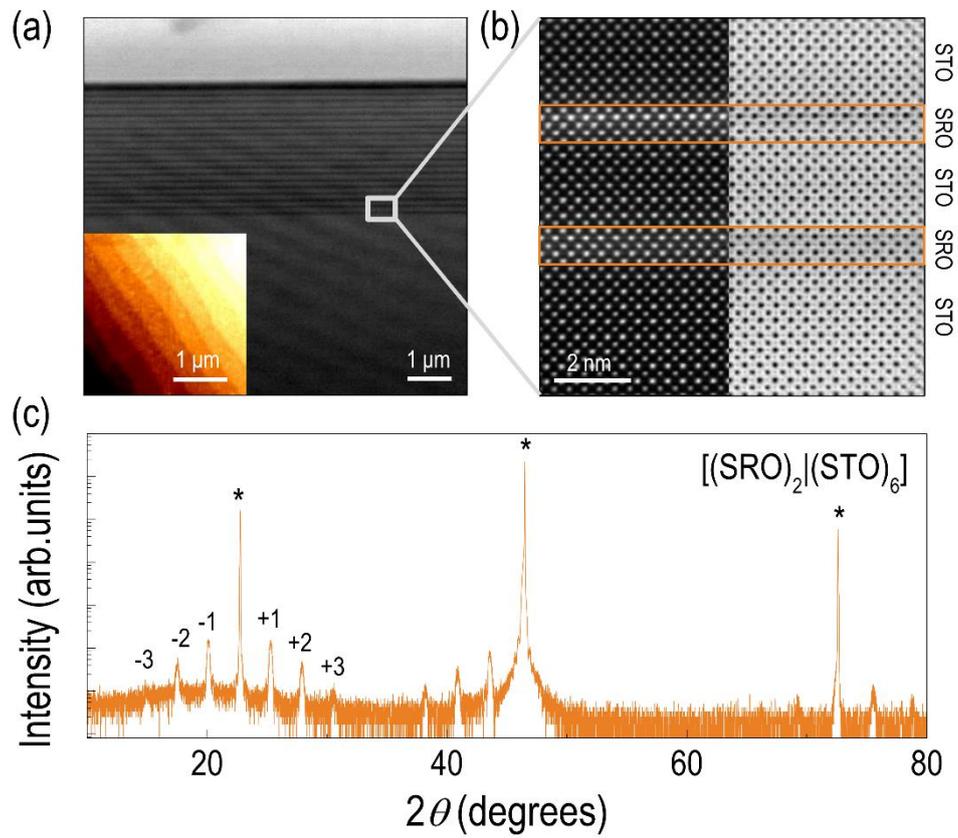

FIG. S2. SRO crystals with customized dimensions. (a) HAADF-STEM images for the $[(SRO)_2|(STO)_6]_{20}$ superlattice (SL) grown on STO (001) substrate in low magnification. Scale bar, 1 μm. The inset shows the surface topography of the SL with an atomically flat step-and-terrace structure measured using atomic force microscopy. Scale bar, 1 μm. (b) HAADF (left side) and ABF (right side) STEM images in high magnification showing the well-ordered atomic columns. ABF-STEM observation also indicates the well-defined epitaxy of the $[(SRO)_2|(STO)_6]$ SLs by clearly visualizing the oxygen atoms. Orange squares indicate the SRO layers. Scale bar, 2 nm. (c) HRXRD $\theta$-$2\theta$ scan indicates the well-defined SLs peaks along with the substrate peaks (marked by *).

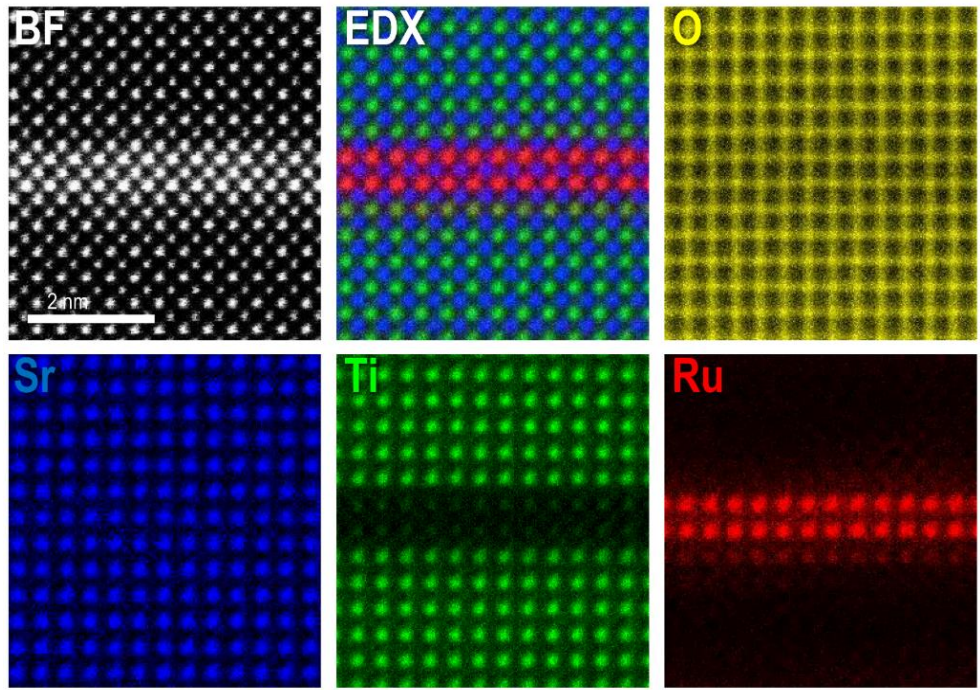

FIG. S3. Atomic-scale STEM-EDX chemical mapping. STEM energy-dispersive X-ray spectroscopy (STEM-EDX) analysis was used to resolve the atomic positions of all the elements in the [(SRO)$_2$|(STO)$_6$] SL, namely, O (yellow), Sr (blue), Ti (green) and Ru (red). All the elements are chemically well-defined at their expected atomic positions.

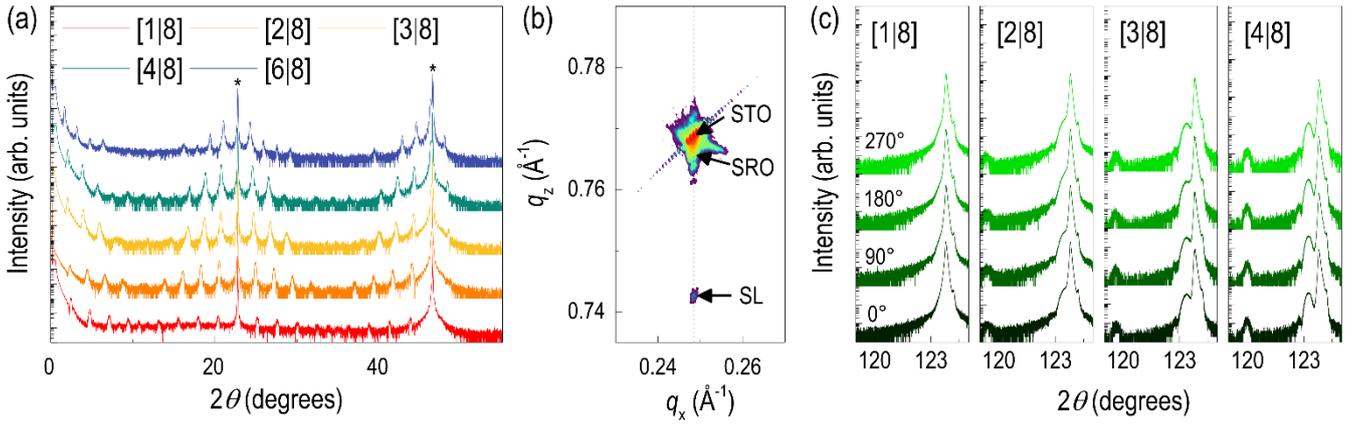

FIG. S4. XRD study of epitaxial $[(SRO)_x|(STO)_8]$ SLs. (a) XRD $\theta$-$2\theta$ scans of different $[(SRO)_x|(STO)_8]$ SL samples. Superlattice peaks corresponding to the supercells of $[(SRO)_x|(STO)_8]$ SLs are clearly visible. The STO substrate peaks are marked by star symbols. (b) Typical reciprocal space map of the SLs, shown for the $[(SRO)_2|(STO)_8]$ SL around the (103) Bragg reflection of the STO substrate, indicating the fully strained state of the SLs with the coherent in-plane lattice constant as that of the substrates. (c) Off-axis XRD scans for the $[(SRO)_x|(STO)_8]$ SLs around the STO (204) Bragg reflections with $\varphi$ angles of 0, 90, 180, and 270°. These results suggest that the SLs with small $x$ have the tetragonal structures without orthorhombic distortions.

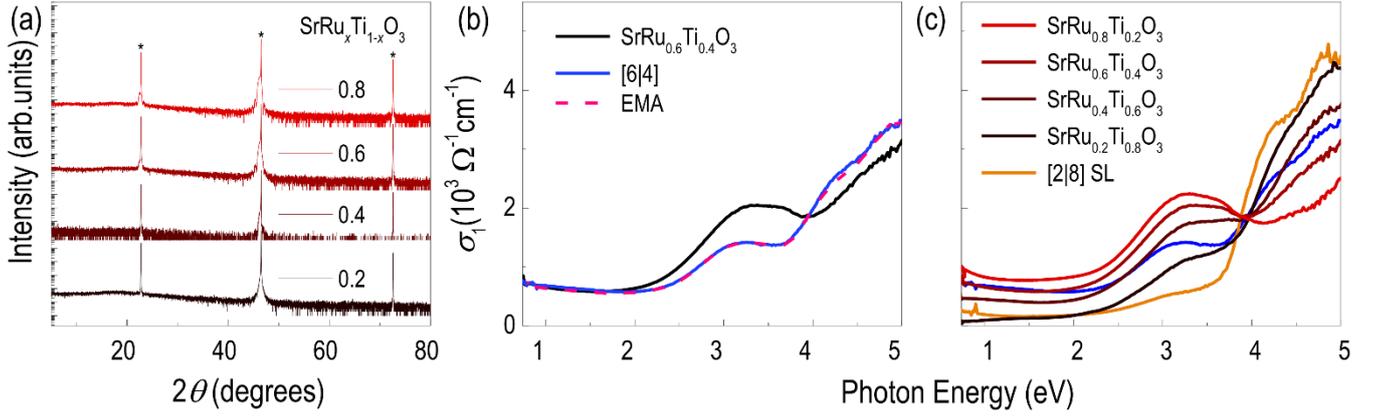

FIG. S5.   Electronic structure of $SrRu_xTi_{1-x}O_3$ solid solution epitaxial thin films. To emphasize the absence of the electronic reconstruction in the layered system of [SRO/STO] SL, we synthesized solid solution epitaxial thin films of $SrRu_xTi_{1-x}O_3$ with $x = 0.1$, 0.2, 0.3, and 0.4. (a) XRD $\theta$-$2\theta$ scans show the structural properties of the high-quality $SrRu_xTi_{1-x}O_3$ solid solution epitaxial thin films. (b) The black and blue solid lines indicate the $\sigma_1(\omega)$ of the solid solution and SL, respectively, with the same SRO volume ratio ($x$ / $x + y$). The pink dotted line was drawn using the 2D effective medium approximation (EMA) (Supplemental Note 3). (c) Solid solution thin films show a clear distinction in $\sigma_1(\omega)$ from those of the SLs.

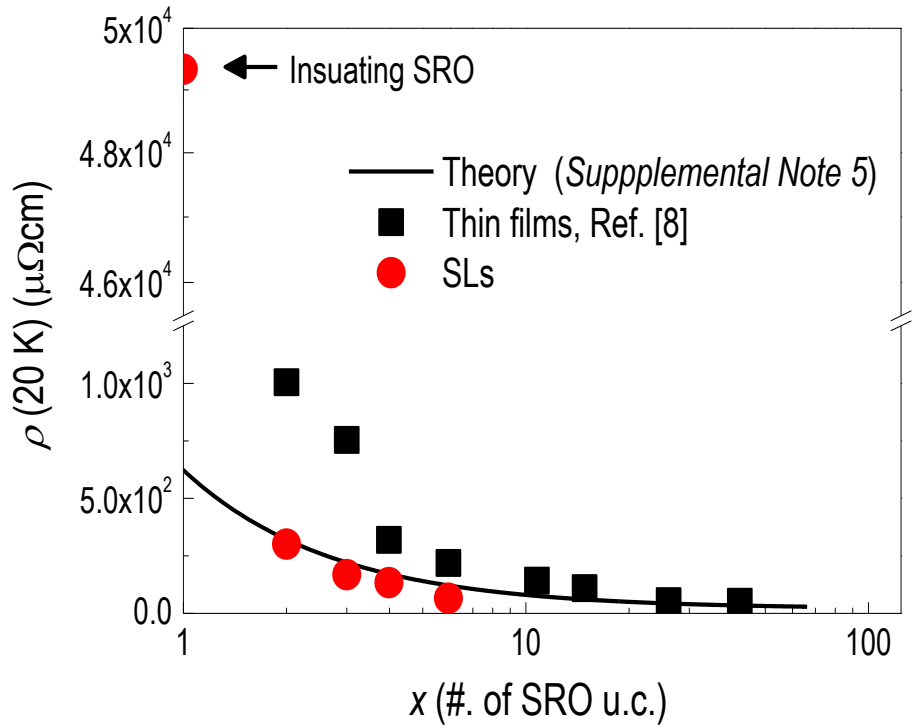

FIG. S6. Thickness-dependent change in $\rho$(20 K). The solid line is the theoretical prediction by the classical Matthiessen's rule [S6] (Supplemental Note 5). The thickness- ($x$-) dependent $\rho$ values at 20 K were plotted for the SLs and thin films for SRO. The black squares indicate the $\rho$ values of SRO thin films from a previous study [8]. The red circles indicate the $\rho$ values of the SLs. On the whole, the $\rho$(20 K)s of the SLs show lower resistivity compared to those of the thin films. They also show excellent agreement with the theory, except for the $x$ = 1 SL, which is insulating.

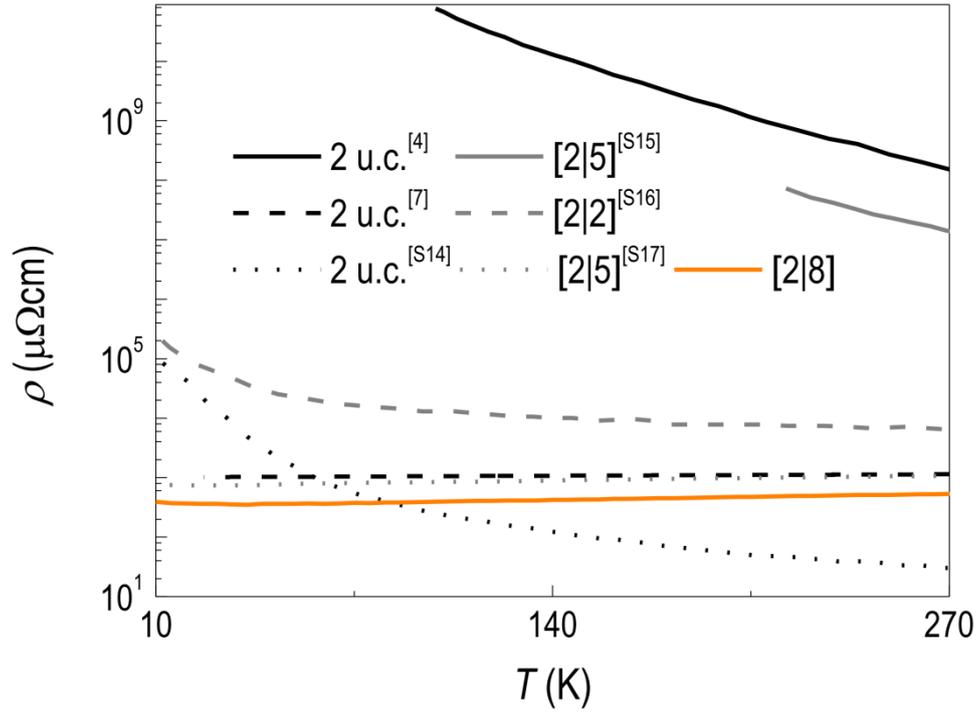

FIG. S7.  Comparison of $\rho(T)$ curves of 2 u.c. of SRO among different SRO systems. Reported $\rho(T)$ curves of SRO thin films and SRO-based SLs with 2 u.c. of SRO thickness are shown. The black solid [4], dashed [8], and dotted [S14] lines indicate the 2 u.c. of SRO thin films, respectively. The gray solid [S15], dashed [S16], and dotted [S17] lines are from the [(SRO)$_2$|(BaTiO$_3$)$_5$], [(SRO)$_2$|(LaAlO$_3$)$_2$], and [(SRO)$_2$|(STO)$_5$] SLs, respectively.

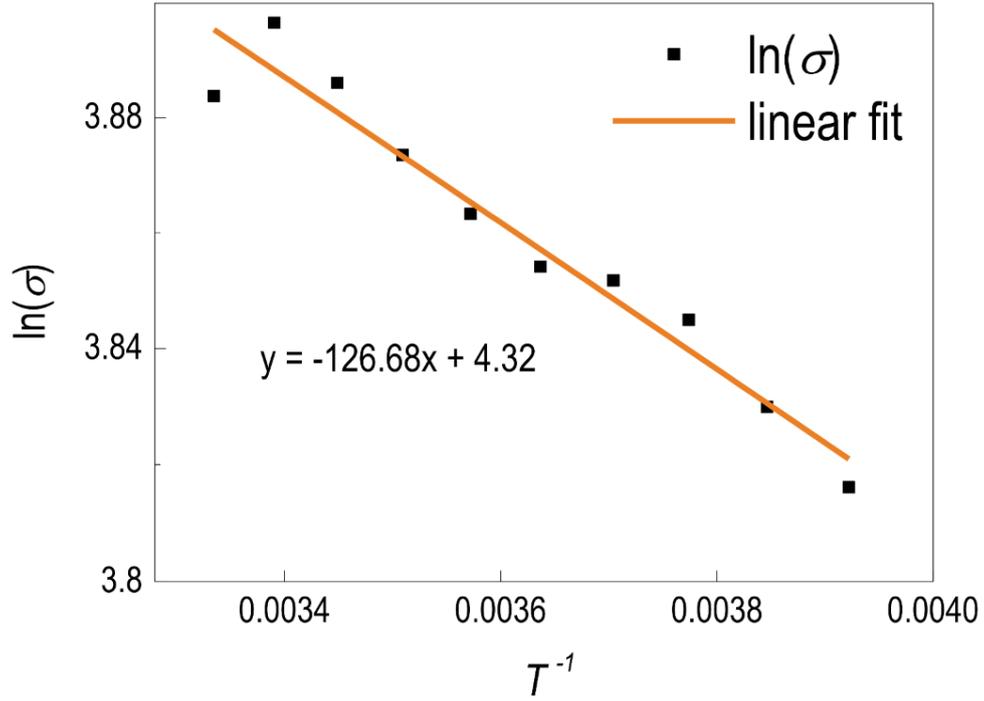

FIG. S8. Estimation of the activation gap of [(SRO)$_1$|(STO)$_8$] SL. We characterized the activation gap of insulating [(SRO)$_1$|(STO)$_8$] SL using the Arrhenius relationship. We used $E_g = 2\alpha k_B / e$ for estimating the activation gap, where $k_B$, $e$, and $\alpha$ are the Boltzmann constant, electronic charge, and linear fitting parameter $\alpha$ from the slope of the ln($\sigma$) vs. $T^{-1}$ plot around room temperature (see inset). The calculated gap value of [(SRO)$_1$|(STO)$_8$] SL is ~20 meV, similar to the optical gap.

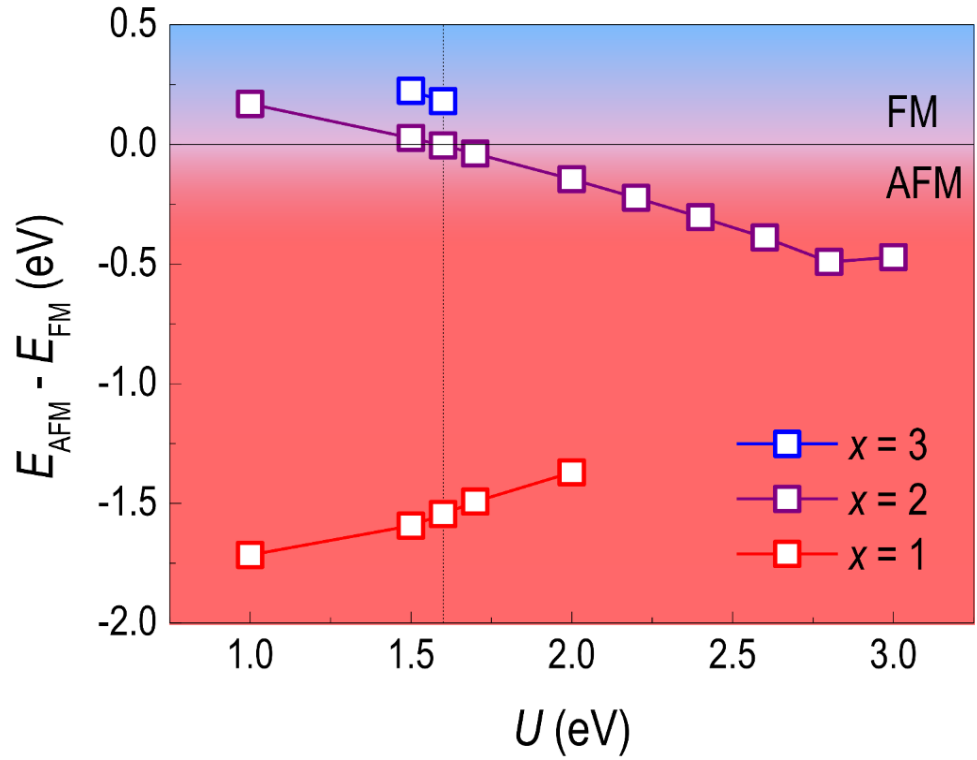

FIG. S9.  Ground state of the SRO layers depending on the on-site Coulomb interaction $U$. Magnetic and electronic ground states of SRO layers are highly dependent on the on-site Coulomb interaction ($U$). The Coulomb interaction of 1.6 eV reproduces the FM-M state for 3 u.c. of the SRO layer, and the G-type AFM-I state for 1 u.c. of SRO layer, which is in good agreement with the experimental observation. For the 2 u.c. SRO layer, the G-type AFM-I and FM-M states are almost degenerate. That is, the AFM-I state is only about 5 meV/f.u. lower than the FM-M state. The red, purple, and blue empty squares indicate the energy differences between G-type AFM-I and the FM-M states for 1, 2, and 3 u.c. of SRO layers, respectively, with the change in $U$.

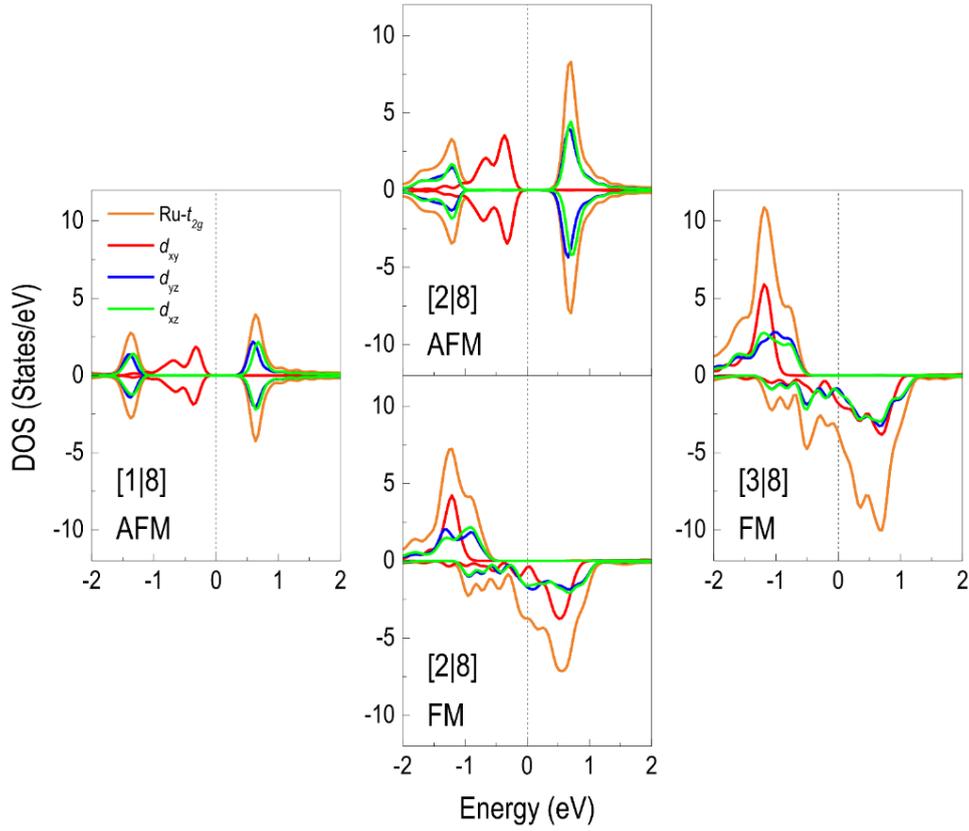

FIG. S10. Orbital selective partial density of states (PDOS) for [(SRO)$_x$|(STO)$_8$] SLs. Up to $x = 2$, SLs have an AFM-I phase as the ground state with a band gap opening between $d_{xy}$ and $d_{yz,zx}$ of Ru-$t_{2g}$ orbitals. On the other hand, $x = 3$ SL has a stable FM-M phase. The electronic structure changes simultaneously from an AFM-I to an FM-M phase depending on the magnetic phases. A dimensionally unstable phase is demonstrated for the $x = 2$ SL, which has a negligible energy difference between AFM-I and FM-M phases. The orange solid lines are the DOS from the total Ru-$t_{2g}$ orbitals. Red, blue, and green solid lines are the DOS from $d_{xy}$, $d_{yz}$, and $d_{xz}$ orbitals, respectively. Vertical dotted lines indicate the Fermi energy.

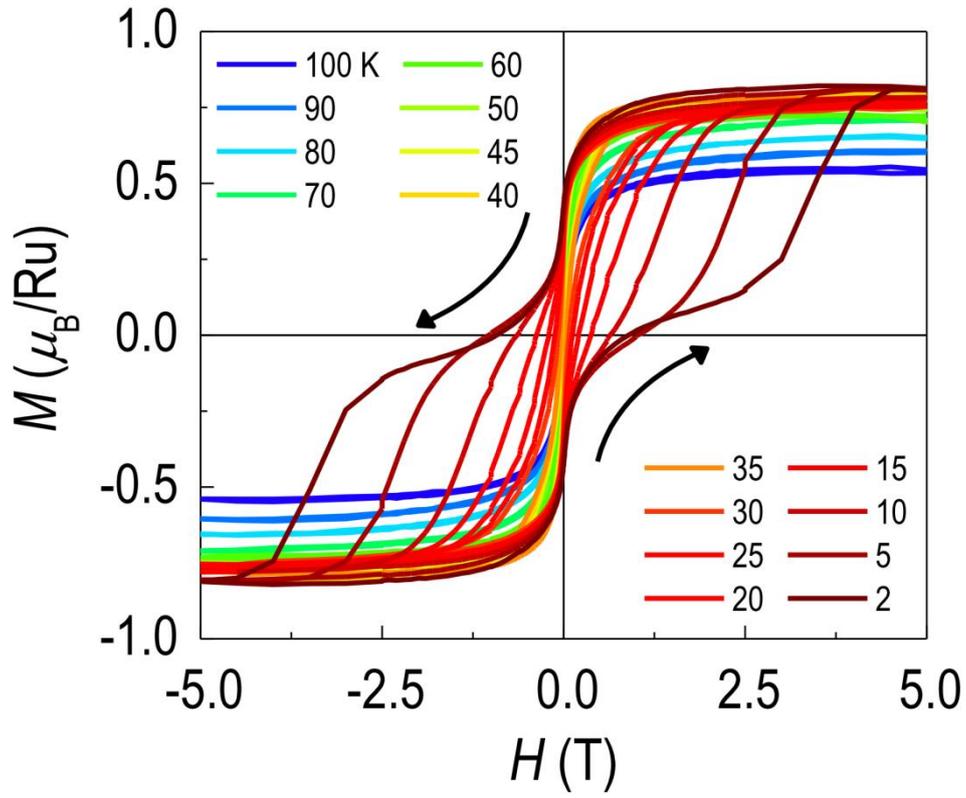

FIG. S11. *M(H)* curves of [(SRO)₂|(STO)₈] SL at different *T*. *M(H)* shows a magnetic phase transition from a FM to an AFM state with respect to *T*, which is characterized by the appearance of a double hysteresis loop below the $T^*$. While the *M(H)* curves below $T_C$ show a typical ferromagnetic single hysteresis loop, they evolve into a double hysteresis loop as the temperature is further lowered below $T^*$. Magnetic field are applied along the out-of-plane direction of the thin film. The arrows of the inset indicate the field directions.

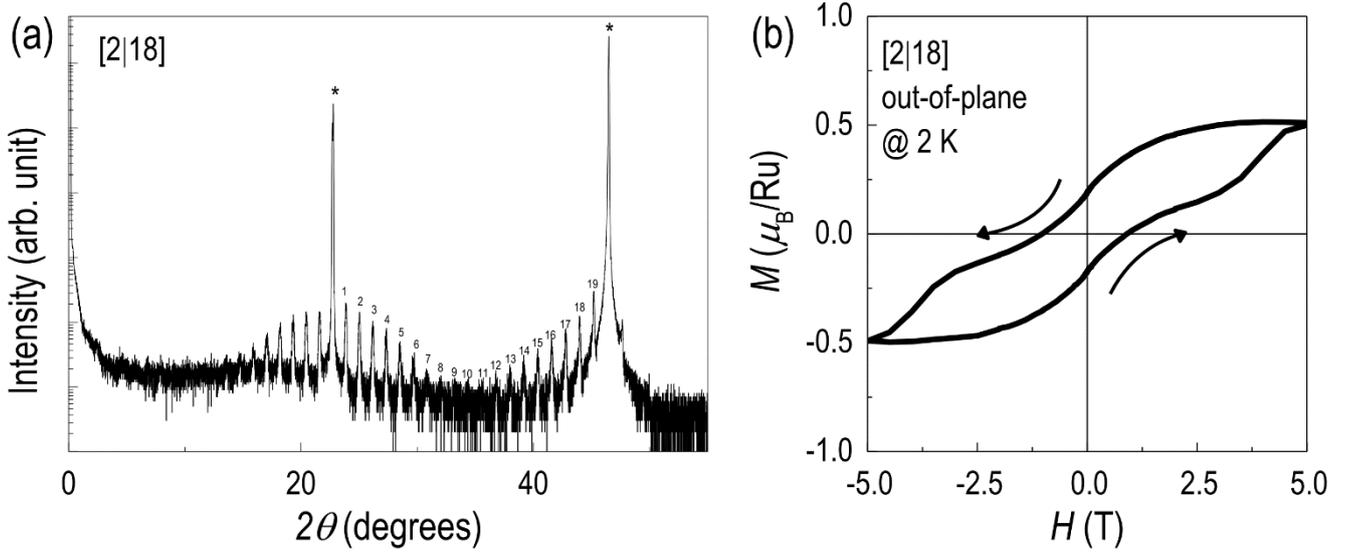

FIG. S12. Structural and magnetic properties of [(SRO)$_2$|(STO)$_{18}$] SL. To exclude the interlayer exchange coupling and exchange bias effect as possible mechanisms of the double hysteresis loop in $M(H)$ of $x = 2$ SL, we synthesized [(SRO)$_2$|(STO)$_{18}$] SL, which has a sufficiently large STO spacer thickness to suppress the interlayer exchange coupling. (a) XRD $\theta$-$2\theta$ scans show sharp superlattice peaks corresponding to the supercells of [(SRO)$_2$|(STO)$_{18}$] SL. The STO substrate peaks are marked by star symbols. (b) $M(H)$ curve measured at 2 K with magnetic field along the out-of-plane direction of the films exhibits a double hysteresis loop, similar to the [(SRO)$_2$|(STO)$_8$] SL. Furthermore, the $M(H)$ curves do not exhibit a meaningful signature of exchange bias [29], eliminating the possibility of the realization of AFM/FM or FM/FM heterostructures in the SL. arrows of the inset indicate the field directions.

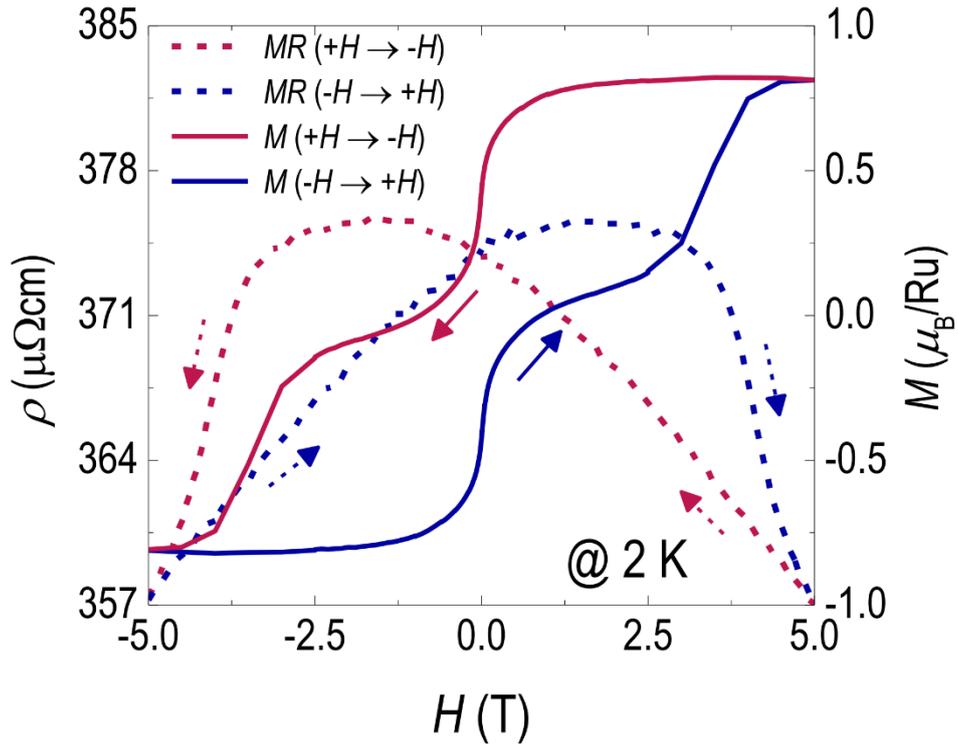

FIG. S13.   MR measurements at 2 K on the [(SRO)₂|(STO)₈] SL. MR measurements provide evidence for a metamagnetic transition. Note that the current and magnetic field are applied along the in-plane and out-of-plane direction of the SL, respectively, for the measurements. The solid (dashed) line indicates the $M(H)$ (MR) curve. The arrows indicate the magnetic field direction. The broad maximum in the MR curve suggests a metamagnetic transition from AFM to FM.